# Evidence for Primal *sp²* Defects at the Diamond Surface: Candidates for Electron Trapping and Noise Sources

Alastair Stacey[1,2]*, Nikolai Dontschuk[1], Jyh-Pin Chou[3]†, David A. Broadway[1], Alex Schenk[5,6], Michael J. Sear[5], Jean-Philippe Tetienne[1,7], Alon Hoffman[8], Steven Prawer[7], Chris I. Pakes[5], Anton Tadich[9], Nathalie P. de Leon[10], Adam Gali[3,4,11], Lloyd C. L. Hollenberg[1,7]

[1]Centre for Quantum Computation and Communication Technology, School of Physics, University of Melbourne, Parkville, VIC 3010, Australia
[2]Melbourne Centre for Nanofabrication, Clayton, VIC 3168, Australia
[3]Institute for Solid State Physics and Optics, Wigner Research Centre for Physics, Hungarian Academy of Sciences, Budapest, POB49, H-1525, Hungary
[4]Present Address: Department of Mechanical and Biomedical Engineering, City University of Hong Kong, Hong Kong 999077, China
[5] Department of Chemistry and Physics, La Trobe University, Bundoora, Victoria 3086, Australia
[6]Present Address: Department of Physics, Norwegian University of Science and Technology (NTNU), N-7491 Trondheim, Norway
[7]School of Physics, University of Melbourne, Parkville, VIC 3010, Australia
[8]Schulich Faculty of Chemistry, Israel Institute of Technology, Technion, Haifa 32000, Israel
[9]Australian Synchrotron, Clayton VIC 3168, Australia
[10]Department of Electrical Engineering, Princeton University, Princeton, NJ, USA
[11]Department of Atomic Physics, Budapest University of Technology and Economics, Budafoki út 8, H-1111, Budapest, Hungary

\* E-mail: alastair.stacey@unimelb.edu.au
† E-mail: jpchou@cityu.edu.hk

**Diamond materials are central to an increasing range of advanced technological demonstrations, from high power electronics, to nano-scale quantum bio-imaging with unprecedented sensitivity.[1] However, the full exploitation of diamond for these applications is often limited by the uncontrolled nature of the diamond material surface, which suffers from Fermi-level pinning and hosts a significant density of electro-magnetic noise sources.[2] These issues occur despite the oxide-free and air-stable nature of the diamond crystal surface, which should be an ideal candidate for functionalization and chemical-engineering. In this work we reveal a family of previously unidentified and near-ubiquitous primal surface defects which we assign to differently reconstructed surface vacancies. The density of these defects is quantified with X-ray absorption spectroscopy, their energy structures are elucidated by *ab initio* calculations, and their effect on near-surface quantum probes is measured directly. Subsequent ab-initio calculation of band-bending from these defects suggest they are the source of Fermi-level pinning at most diamond surfaces. Finally, an investigation is conducted on a broad range of post-growth surface treatments and concludes that none of them can reproducibly reduce this defect density below the Fermi-pinning threshold, making this defect a prime candidate as the source for decoherence-limiting noise in near-surface quantum probes.**

The quantum properties of the negatively charged nitrogen-vacancy (NV) defect centre in diamond [3] have enabled a remarkable range of nanoscale sensing demonstrations. Its capabilities include sensitivity to magnetic,[4] electric,[5] strain[6] and thermal[7] fields, with demonstrated applications ranging from condensed matter physics[8, 9] to biology.[1, 10] In many of these applications the NV centre must be simultaneously in a negative charge state and located close to the crystal surface, to enable sensing of extrinsic targets. Satisfying these conditions without deleteriously affecting the quantum coherence (and sensitivity) of the NV sensor is a significant challenge,[2, 11] with parallels to the '$1/f$' noise problem found common to quantum system surfaces.[12] There have been a range of studies attempting to understand both the nature and causes of NV decoherence at the diamond surface, including magnetic noise spectroscopy[2, 13] and various surface chemical treatment protocols.[1, 14, 15] Recent studies have suggested that the presence of simple paramagnetic bath noise is insufficient to explain the decoherence characteristics of near-surface NV centres, and that some form of phonon-related interaction is required.[2, 16] Other experiments have demonstrated that electric field noise is an important factor, affecting both the decoherence[17] and longitudinal relaxation properties[18] of the NV spin system. The need for a negative charge on the NV centre imposes additional requirements on the treatment of the sample surface,[19, 20] including the removal of diamond's natural hydrogen termination,[21] because this termination induces a sub-surface hole accumulation layer that favours the $NV^0$ rather than the $NV^-$ state. This is typically dealt with by oxidative treatments, however NV charge instabilities still remain for very shallow NVs, manifesting as decreased $NV^-/NV^0$ ratios or reduced contrast in optically detected spin-state measurements[22], where the charge state of the NV is dynamically modified (in thermal non-equilibrium) by photons during these experiments.[23]

There is no unifying explanation for the different levels of quantum coherence and photophysical problems in near-surface NV centres. However, surface electrons and traps are inherently implicated as sources of both electric and magnetic fields due to individual trapped electrons having a guaranteed non-zero magnetic moment in their naturally unpaired state. NV-based studies have estimated the surface spin densities to be of order $1 \times 10^{13}$ cm$^{-2}$,[2, 13, 14] which is surprisingly similar to the apparently ubiquitous surface charge density seen at a variety semiconductor and superconductor surfaces [24], despite the lack of a detailed understanding of the chemical and crystal origin of these surface defects.
In diamond, Fermi-level pinning is also often observed[25] without a robust microscopic explanation. An interplay between hydrogen and oxygen termination states are suspected,[22] but no detailed analysis exists that supports the contention that insufficient surface oxygen bonding is the cause of electron trapping, beyond the heuristic that higher positive electron affinity is needed to increase the Fermi-level within the diamond.[20, 26] A recent theoretical study has reported the likely effects of different surface states on NV photophysical properties,[27] however these effects are short-range (<5 nm) and no unoccupied electronic states below the $NV^-/NV^0$ transition energy have been identified. Because of the inherent abundance of nitrogen-donors ($N_s$ centres) in any NV-containing crystal there must exist surfaces states with unoccupied electronic levels lower in the band-gap for near-surface NV centres to experience a local Fermi-level which is below the transition level of $E(NV^-/NV^0) \approx 2.7$ eV (above the VBM - valence band maximum) [28]. Microscopic identification of these electron traps is thus of critical importance for both electronic and quantum technologies that use diamond surfaces.

**Protected *sp²* defects in diamond surface vacancies**
With the advent of ultra-pure diamond synthesis and purification technologies (e.g. strong acid boiling and plasma etching[29]) it has been typically assumed that both the diamond bulk and surface are free of any graphitic carbon. Here, we introduce evidence for the presence of stable primal C=C bonds at the diamond surface, which we denote as '*sp²* defects'. These primal C=C bonds are distinguishable from the natural C=C bond formation found on bare reconstructed (001) diamond surfaces[30] in that the primal defects are stable under ambient conditions and their σ bond is localized to a normal tetrahedral diamond bonding position. These primal $sp^2$ defects are geometrically prohibited on perfect low-index diamond surfaces but are likely to populate surface defects, such as surface vacancies. **Figure 1a(middle)** shows an example of such a situation, where one of the surface carbon atoms has been removed from the (001) surface, leaving a surface vacancy. The two sub-surface carbon atoms which were previously bonded to the now-removed surface carbon have dangling bonds which must be satisfied. Due to the small size of this surface vacancy and the stiffness of the diamond lattice, it is difficult to find an atomic species that can fit in this vacancy and satisfy both dangling bonds simultaneously. As such the most natural consequence is that one (**Fig 1a**) or both (**Fig 2b**) second-layer carbons form C=C bonds with their remaining surface-carbon neighbours. The surface vacancy thus acts to protect these primal $sp^2$ defects from chemical reactions and allows for a family of defects, depending on the size and co-termination chemistry of the vacancy. The common element of these $sp^2$ defects is the presence of one or more C=C bonds, and as such we searched for experimental evidence of $sp^2$ hybridised carbon, quantified its areal density and investigated the resultant energy level structure associated with specific defects.

**Measurement of $sp^2$ hybridised carbon at the diamond surface**
There are few measurement techniques that can identify the phase composition of carbon atoms with a sub-monolayer coverage on a diamond surface. However, 55° (sample/beam angle) near-edge x-ray absorption fine-structure spectroscopy (NEXAFS) at the C1s-edge is well suited to this analysis, as it is equally sensitive to $sp^2$ and $sp^3$ hybridised carbon, and the characteristic π* (285 eV) and σ* (>288 eV) resonances are well separated in energy.[31] In fact, the application of this technique to typical diamond surfaces is well known to show the presence of C=C carbon at diamond surfaces,[32] to the extent that samples are typically assumed to either have 'carbon contamination',[33] and/or there is a presumed problem with the normalization technique.[34] Although carbon contamination in the synchrotron beam-line can act to compromise the quantitative reliability of the data collected, there are known effective methods for the removal of these contaminant signals, such as double-normalization,[35] and *in-situ* heating (>400 °C) for the physical desorption of physisorbed adventitious carbon. After utilization of these techniques on dozens of different samples, the only samples repeatedly observed by us with very low C=C signatures (close to the technical noise) are those that have been recently grown and are inserted into the measurement chamber in an as-grown (hydrogen terminated) state. However, on all other surfaces, after *in-situ* removal of physisorbed species, a significant signal of C=C at 285eV is observed, along with associated changes to the σ* spectral region (see Supp. Info.). In Figure 1d a selection of pre-edge spectra are displayed. These consist of spectra from diamonds which are as-grown and consecutively acid-boiled (sulphuric acid and sodium nitrate) and then $O_2$-burned at 465°C, with data displayed after double-normalization.[36] These sample treatments are directly relevant for comparison with near-surface NV reports and can be used to directly probe the validity of the 'selective' burning of $sp^2$ carbon over $sp^3$ carbon at 465°C.[19].

Quantitative estimation of the surface coverage of C=C can be conducted with this data,[37] although this is complicated by the inhomogeneous character of the surface-only layer. In order to overcome this issue we have utilized a graphene-on-diamond sample[8] as a full mono-layer C=C coverage reference. The surface coverage estimate is thus conducted using the surface fraction

$$f_{\text{C=C}} = \frac{I_{\text{sam}}^{\pi*}/I_{\text{sam}}(\Delta E)}{I_{\text{ref}}^{\pi*}/I_{\text{ref}}(\Delta E)} \qquad (1)$$

where $I_{\text{sam}}^{\pi*}$ represents the fitted peak-area of the C=C resonance at 285eV, and $I_{\text{sam}}(\Delta E)$ represents the integration of the σ* signal between 288.6eV and 320eV. It should be noted here that a coverage fraction of $f_{C=C}$ = 100% indicates a coverage equal to the full monolayer (ML) of the graphene reference sample ($\approx 3.8 \times 10^{15}\ cm^{-2}$), which has a somewhat higher atomic areal density than the (001) surface of diamond ($\approx 1.6 \times 10^{15}\ cm^{-2}$). The presence of a thin layer of intercalated adventitious carbon and/or water between the graphene and diamond layer could decrease $I_{ref}(\Delta E)$, thus producing an over-estimation of the $f_{C=C}$ value (but not more than a factor of about 2x – see Supp. Info.). On freshly as-grown (<24hrs old) hydrogen terminated diamond samples this method yields a fractional C=C coverage below 1%, which becomes hard to distinguish from technical noise. The same procedure was followed for a variety of other surface modification techniques, including Ozone/UV[22] and SF6 plasma treated[38] samples, and these values are included in the C=C estimates in Table 1. We note that the lowest C=C density (1.6%) is achieved by a combination of wet and dry treatments (Acid + $O_2$ Burn + Piranha), similar to the process used in Ref. [1] and observed to result in relatively long near-surface NV coherence times. Although, a larger C=C fraction (>3%) was found in another sample following this treatment process.

**Filling of surface *sp²* defect states: band-bending and trapped surface electrons**
While there is some variability in the C=C content between surface treatments, and the selective $O_2$ burning process does indeed reduce the C=C concentration, as predicted,[19] no samples (other than as-grown) were observed with a defect density below 1% of a graphene monolayer, which represents more than $4 \times 10^{13}$ defects/cm². This is a higher areal density than the observed free-electron density at diamond surfaces,[2, 13, 14] making surface C=C bonds a good candidate for the microscopic origin of these electron traps, as well as charge-state stability issues with near-surface NV centres.[22]

Modelling the effect of these surface *sp²* defects on NV and electronic devices requires an understanding of their ability to trap free electrons, which in turn requires first that their unoccupied energy levels are understood. The conduction band minimum is directly observed in the NEXAFS spectra (at 289eV), and the π* unoccupied state resonance is seen at 285eV, suggesting that the lowest unoccupied level from these defects will be $E_{C=C} \approx$ 1.5eV above the valence band maximum. However the exact energy resonance associated with NEXAFS measurements, being core-level optical excitations, is modified by final-state effects [39], meaning that this value cannot be directly used as an estimate of the true *sp²* defect density of states. As a result, we turn to *ab-initio* calculations to provide a more direct estimate of this electron trap energy level.

The calculations were performed with the Vienna *Ab-initio* Simulation Package (VASP) using the projector augmented-wave (PAW) method[40] to represent the electron-ionic core interactions. The gradient-corrected Perdew-Burke-Ernzerhof (PBE) functional[41] was implemented for exchange and correlation estimation, while the screened hybrid functional HSE06[42] was used for final electronic energy level determination, as it provides accurate electronic properties of diamond, including bandgap and surface bands[28] (see Sup. Info. for more detail). To investigate the plausibility of the proposed *sp²* defect structure from Figure 1, using a C(100)-6×6-14L surface model, we examined more than one hundred model variations, including different numbers of surface vacancies, different co-termination species and configurations, and isolated or paired C=C bonds, and accounted for relaxation energies for the unoccupied defect levels and size effects (see Sup. Info.). The acceptor level of *sp²* defect states in most models appears at 2.2 eV above the valence band maximum (VBM), while the acceptor level of the two-aligned C=C model (**Figure 2a**) produces an acceptor level at 1.78eV above the VBM, likely due to interaction between the C=C orbitals. The relevant relaxed lattice parameters are shown in this figure, with the lengths of C-C and C=C bonds sitting in the normal ranges around 1.39 Å and 1.52 Å respectively. The energy spread of 0.4eV from these models is smaller than the normal inhomogeneous π* peak width found in the NEXAFS measurements (0.6-0.8eV), and as such a variety of *sp²* defect models, including single and paired C=C bonds and different co-terminations, are likely to exist on any real-world sample surface.

It should be noted that this C=C bond configuration is also possible on any intersection between a (001) and (111) surface plane and with any chemical termination, even at the nanoscale, and as such these results are likely extensible to other real-world low-index single crystal surfaces as well as nanocrystalline diamond applications.

As the *sp²* defects introduce unoccupied electronic levels in the range 1.5-2.2eV above the valence band minimum, and all NV-containing samples contain n-type dopants ($N_s$ centres) in the bulk, these surface defects are likely to be at least partially filled and result in upward band bending toward the diamond surface. The areal density of trapped electrons ($Q_{C=C}$) can be calculated simply via the filling factor of the surface defects, as a function of the defect concentration ($N_{C=C}$), defect energy level ($E_{C=C}$), magnitude of the band-bending ($V_{bb}$) and height of the bulk Fermi level ($E_b$):

$$Q_{C=C} = \frac{eN_{C=C}}{\exp\left(\frac{E_{C=C} + V_{bb} - E_b}{k_B T}\right) + 1} \tag{2}$$

Following the treatment of Mönch,[43] for a system such as this with static depleted sub-surface impurities and no significant minority carriers, the areal density of depleted bulk donors (space-charge), can be found by application of the extrinsic Debye length (with $\varepsilon_d = 5.7$), bulk donor density ($N_d$), and band-bending ($V_{bb}$), with approximate analytical solution:

$$Q_{sc} \approx \sqrt{2\varepsilon_0 \varepsilon_d N_d k_B T [e^{-|V_{bb}|/k_B T} + V_{bb}/k_B T - 1]} \tag{3}$$

The system will thus be in equilibrium when the surface and space charge concentrations are in equilibrium, leading to a solution for the band-bending height $V_{bb}$. For the case of our N-implanted diamond, the substitutional nitrogen donors will pin the *bulk* Fermi level (we estimate this as $E_b = 4.6$eV above the VBM – see Supp Info.). The bulk dopant density $N_d$ can be estimated as $N_d = 1 \times 10^{15}$ cm$^{-3}$. This represents a bulk nitrogen donor concentration of only ≈10ppb which is reasonably close to the lowest achieved values in diamond growth, and corresponds to the density obtained by ion implantation in our experiments (see details below). This uniform dopant concentration then allows for a linear-approximation of the band-bending depletion depth, by assuming all donors in the band-bending region are ionized, and through application of the Poisson equation. This gives a total band-bending depth of:

$$d = \sqrt{\frac{2\varepsilon_0 \varepsilon_d V_{bb}}{eN_d}} \tag{4}$$

Here we will use the lowest density functional theory derived acceptor level of $E_{C=C} = 1.78$eV (above the VBM), however there is little qualitative difference in band-bending for a range of acceptor values between 1.5 - 2.2eV.

A numerical calculation for our NV sample is shown in **Figure 2d**, for a range of C=C coverages, including a conservative estimate of C=C surface coverage at 0.1% of the graphene monolayer reference. In this case the expected band-bending extends approximately 1,000 nm into the sample, with the Fermi level kept below the NV⁻/NV⁰ charge state transition[28] within 250 nm of the surface. Even in this simplified model, with a semi-infinite donor layer ($N_d = 1 \times 10^{15} N/cm^3$), the ionization of surface acceptors sites ($Q_{ss} \sim 1 \times 10^{11} e/cm^2$) is essentially limited by the donor density. An increase in donor density will thus increase the $Q_{ss}$ value without significantly affecting the band-bending shape and surface Fermi level position. This 'Fermi-level pinning' is characteristic of surface defects at semiconductor interfaces, and even in this low-damage example would require an implant of more than $1 \times 10^{12} N/cm^2$ to overcome. Re-application of these models for different C=C densities shows that a surface coverage of as little as 0.003% will induce band-bending below the NV⁻/NV⁰ transition level in the top 40nm of diamond (Figure 2c).

Thus it can be easily understood that surface *sp²* defects under any of the currently utilized surface treatment (and termination) methods will fill with between $10^{11}/cm^2$ and $10^{13}/cm^2$ trapped charges, and deplete the 'dark' charge state of dilute NV centres near the sample surface. This also provides a viable explanation for a variety of Fermi-level pinning predictions and observations of Schottky barrier heights in similar diamond surfaces.[25]

**Effect of *sp²* defects on NV charge state dynamics and quantum coherence**
The apparently contradictory observation of NV centres in their negative charge state near these surfaces is not well described by this thermodynamic equilibrium analysis because the NVs are only observed under intense photo-illumination. It should also be understood that the almost complete lack of intrinsic carriers in diamond means that all of the charges in the band-bending region are immobile, rendering an homogeneous Fermi level assumption somewhat problematic.[44]

To monitor the impact of changes in the *sp²* defect content on near-surface NV properties, we have produced an NV-implanted sample, implanted with $1 \times 10^9$ ¹⁵$N^+/cm^2$ at 3.5 keV ion energy, and annealed at 950 °C for 3 hours in vacuum. These samples are thus expected to comprise single NVs with depth distribution approximately 5-15 nm,[45] for which we can apply standard quantum metrology methods. To investigate dynamic NV charge state issues the Rabi-flopping protocol has been used, as it is sensitive to the charge-state of the NV centre in the absence of photo-illumination.[46] In this way, NV centres that lose their negative charge state during ground-state evolution, due to the thermal-equilibrium Fermi level being below the NV⁻/NV⁰ transition level, are not able to respond to the ground-state microwave driving and show up as having significantly reduced 'Rabi' contrasts. Here the contrast is normalized to the intensity of the polarized 'bright state', and hence we define contrast as

$$\text{Contrast} = \frac{F_{\text{top}} - F_{\text{bottom}}}{F_{\text{top}}} \qquad (5)$$

where $F_{\text{top}}$ ($F_{\text{bottom}}$) represents the fluorescence rate of the top (bottom) of the Rabi oscillation curve. Bulk NV centres are typically observed to give Rabi contrasts around 30% although it should be noted that the measurement parameters (integration times and intensities) can change this value in the 20-40% range for bulk defects.

A total of 45 individual (randomly selected) NV centres were studied, and the Rabi contrast of each NV was measured before and after selective O₂ burning. An example of Rabi oscillations measured for a representative single NV is shown in Figure 3b, and the full data set of extracted Rabi contrasts is displayed as a histogram in **Figure 3c**, where it can be clearly seen that a dramatic improvement in Rabi contrast is observed for all NV centres upon O₂ burning, consistent with a similar study recently conducted on selectively O₂ burned NV diamonds.[22] It should also be noted that the contrast of O₂ burned NV centres was found to be stable, while the Rabi contrast (and overall fluorescence rate) of acid-boiled NV centres typically reduced with repeated measurements (or increased laser intensity). These results are consistent with the picture that reduced *sp²* defect density at the diamond surface leads to improved NV⁻ dynamic charge state stability, however, the non-equilibrium nature of this experiment (with the application of high intensity pulsed laser illumination) makes quantification of the dynamic surface trapped charge difficult.

While specific signatures of electron spins at the diamond surface can be seen with double electron-electron resonance (DEER) studies [47], there have been reports of near-surface NVs which experience reduced decoherence but produce no observable DEER signature[48] (including the NVs measured in this work). It is worth noting however that if the *sp²* defect sites are the cause of most trapped surface charge, there may be significant mobility of the charge between proximal and high density unoccupied *sp²* defect sites, inhibiting the EPR signature required to observe them with a DEER experiment. Similarly, the relaxation times and characteristic bath noise spectral density generated by such spins, may be affected by such migration, as well as the average distance (coupling) between them, and other effects such as increased spin-orbit interactions due to the *sp²* hybridization.

Our results suggest that most diamond samples will exhibit between $10^{11}/\text{cm}^2$ and $10^{13}/\text{cm}^2$ surface trapped electronic charges, which is in good agreement with the experimentally determined range, from NV decoherence measurements.[2, 13, 14] This indicates that the *sp²* defect is a good candidate to explain the limited coherence of near-surface NV centres, although there are likely to be a combination of effects rather than only one, including the interplay between spin and charge dynamics within the ensemble of *sp²* defects, that explain this decoherence. We note that the coherence times of near-surface NVs produced in this work (e.g., using the acid boiled + O₂ burned process) were comparable to those reported in recent works [1, 49] (see SI).

In conclusion, the apparent Fermi-level pinning and NV charge state properties at diamond surfaces are likely to be dominated by previously unrecognized primal *sp²* defects. These defects are observed in significant densities on all diamond surfaces, with the lowest defect density being measured at $f_{C=C} \approx 1\%$ of a surface monolayer. Surface treatments that reduce this defect density appear to be correlated with improved NV⁻ dynamic charge stabilities, but do not correlate with significant improvement in NV coherence properties. This is consistent with a Fermi-level pinning scenario, where even a 0.1% surface coverage of *sp²* defects pins the surface Fermi level below the NV⁻/NV⁰ transition energy in low-density NV samples.
This motivates further study of the charge state dynamics in NV samples under photo-illumination, and suggests that future optimization of the surface chemistry[27] will also require separate control of the surface primal *sp²* defect density. While these studies were conducted on (001) oriented low-index diamond substrate surfaces, the nature of the *sp²* defect is likely to be extensible to other damaged crystal faces, as well as nanocrystalline diamond applications. Finally, thermodynamic equilibrium simulations suggest that in representative NV-implanted samples the density of trapped (unpaired) electron spins will fall within the range of experimentally measured values, making this *sp²* defect a good candidate for the source of electric and magnetic noise at diamond surfaces. As such, we predict that this defect must be addressed before surface bound electromagnetic noise sources in diamond can be eliminated.

**Experimental Section**

*Sample Preparation:* (001) diamond samples were grown by chemical vapor deposition in a Seki 6300 reactor, using standard methane and hydrogen mixtures, and with trimethylborane as the boron-precursor. Hydrogen termination was conducted separately with a low level of methane <1%, to maximize surface quality. Acid boiling was conducted with a mixture of sodium nitrate and sulphuric acid at a temperature above 240°C for at least 10 minutes, and samples were rinsed with DI water after boiling. $O_2$ burning was conducted in a 'tube-style' oven with flowing pure $O_2$ gas, at a calibrated temperature of 465°C.

*NEXAFS Measurements:* Carbon K-edge NEXAFS measurements were conducted at the Soft X-Ray beamline of the Australian Synchrotron. All measurements were conducted in the partial electron yield (PEY) mode, recorded either with a channeltron or multi-channel plate, with retarding voltage typically set to 220V, to collect only high energy carbon Auger electrons and thus data from the very near-surface region (<1nm) of the samples. Double normalization was conducted with an 'IO' gold grid partially inserted into the beam, to monitor fluctuations in beam intensity for all scans, and normalized to a reference photo-diode scan for beamline transmission function (primarily for removal of beamline carbon-based absorption characteristics). Sample measurements are presented from a variety of different beamtime experiments, with new normalization scans taken during each experimental run.

*NV Measurements:* NV measurements were conducted with a home-built confocal apparatus, capable of a comprehensive set of microwave-driven optically detected magnetic resonance and coherence measurements, including included EPR identification of the $^{15}$N isotopic label, spin-lifetime ($T_1$) and spin-coherence ($T_2$) measurements, as described in Ref [50].

**Acknowledgements**

This work was supported in part by the Australian Research Council (ARC) under the Centre of Excellence scheme (project No. CE110001027). This research was undertaken on the Soft X-Ray Spectroscopy beamline at the Australian Synchrotron, part of ANSTO. This work was performed in part at the Melbourne Centre for Nanofabrication (MCN) in the Victorian Node of the Australian National Fabrication Facility (ANFF). A.G. acknowledges the support from the National Research Development and Innovation Office of Hungary within the Quantum Technology National Excellence Program (project no. 2017-1.2.1-NKP-2017-00001). L.C.L.H. acknowledges the support of an ARC Laureate Fellowship (project No. FL130100119). J.-P.T. acknowledges support from the ARC through the Discovery Early Career Researcher Award scheme (DE170100129) and the University of Melbourne through an Establishment Grant and an Early Career Researcher Grant. D.A.B is supported by an Australian Government Research Training Program Scholarship.

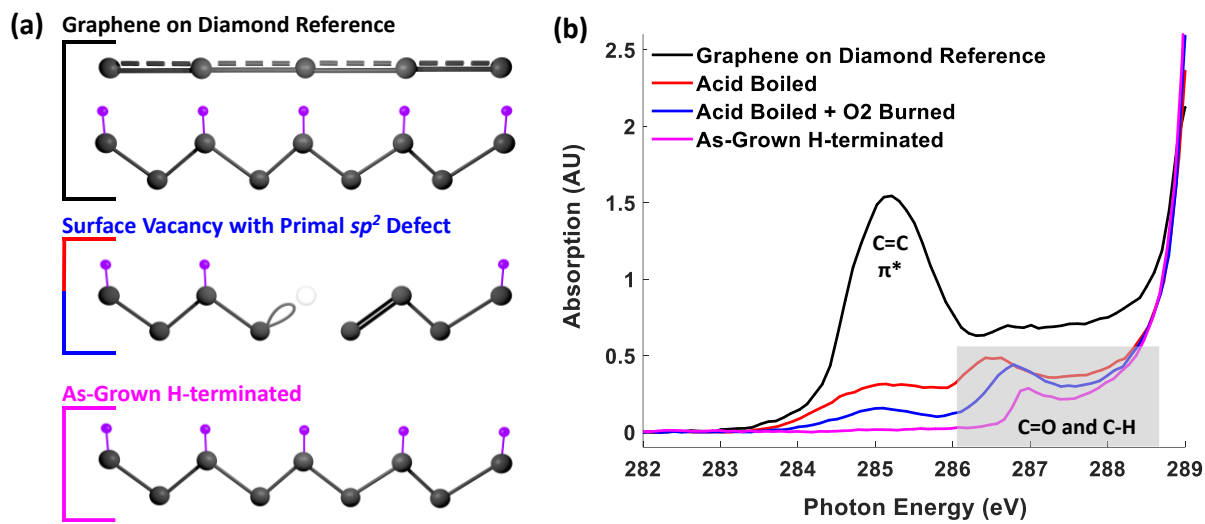

**Figure 1.** (a) Diamond (001) surface models, including a reference as-grown hydrogen terminated surface (bottom), a partially reconstructed surface vacancy with a single primal $sp^2$ defect (middle), and a reference graphene-covered surface (top). (b) C1s pre-edge NEXAFS scans of various diamond surfaces after in-situ cleaning, showing intrinsic $\pi^*$ resonances at 285eV photon energy. These can be compared with a monolayer graphene-on-diamond reference scan (black), enabling the quantification of this surface-bound C=C carbon on all samples. Oxygen and hydrogen related NEXAFS peaks are shown in the greyed region.

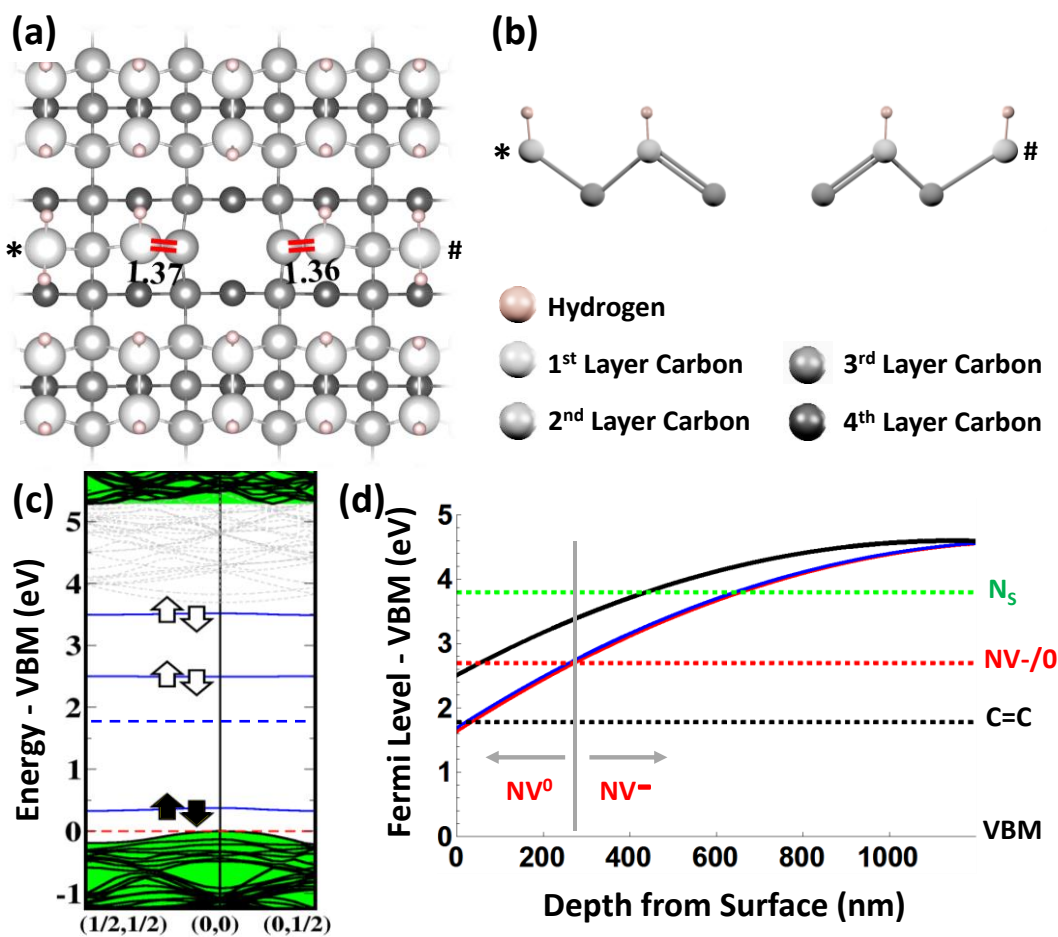

**Figure 2.** Theoretical (density functional theory) relaxed model. (a) A pair of $sp^2$ defects within a single surface vacancy – with bond lengths indicated in Å. (b) A cross-section of the first two layers of carbon. (c) Calculated energy levels of this system, with empty bands in the band-gap shown with open arrows (spin states). Once relaxation energies and size effects are included the lowest lying adiabatic acceptor level is calculated to be at 1.78 eV above the valence band maximum (shown as a dashed blue line). The bulk valence and conduction bands are green coloured whereas the surface bands due to hydrogenation are depicted in faint grey colour. (d) The resulting surface band-bending within a uniformly nitrogen-doped diamond sample is simulated with a bulk nitrogen concentration of 10 ppb and an $sp^2$ defect coverage 0.003% (black), 0.1% (blue) and 1% (red). It can be seen that for coverages above 0.1% the thermal-equilibrium Fermi level position is too low for negatively charged NV centres to be stable within 250 nm of the surface.

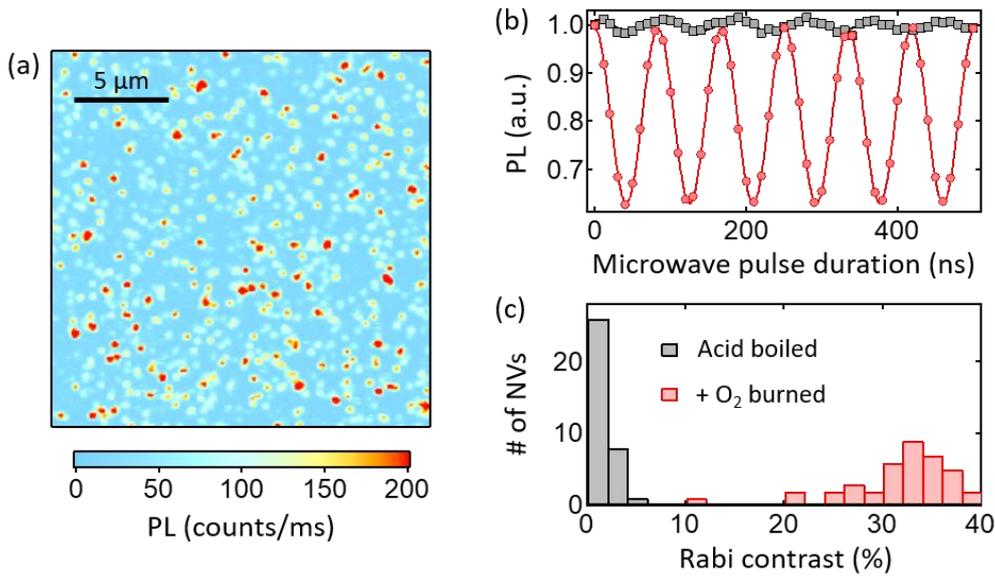

**Figure 3.** (a) Confocal microscope image of a near-surface implanted NV sample (measured depth range 2-12nm). (b) Rabi oscillations measured for a single 8.2nm deep NV after acid boil (black data) and for the same NV after subsequent $O_2$ burning (red). Solid lines are sinusoidal fits. (c) Histograms of the Rabi contrast of 40 single NVs randomly selected in the region displayed in (a) after acid boiling (grey), and of the same NVs measured after subsequent $O_2$ burning (red).

**Table 1.** Estimates of the intrinsic $sp^2$ defect density ($f_{C=C}$) at various diamond surfaces, with a 100% monolayer defined as the C=C density of a monolayer graphene sample. Freshly as-grown samples show a variety of $sp^2$ defect densities below 1%, which is close to the technical noise.

| Sample Treatment | Freshly As-Grown Diamond | Graphene on Diamond | Acid Boiled | Acid + $O_2$ Burned | Acid + $O_2$ Burned + Piranha | $O_2$ Burned (only) | Ozone/UV | SF6 Plasma |
|---|---|---|---|---|---|---|---|---|
| C=C fraction [$f_{C=C}$ %] | <1 | 100 | 18.6 | 7.7 | 1.8 | 6.1 | 3.7 | 4.0 |

# Supplementary Information for the manuscript "Evidence for Primal *sp²* Defects at the Diamond Surface: Candidates for Electron Trapping and Noise Sources"

Alastair Stacey, Nikolai Dontschuk, Jyh-Pin Chou, David A. Broadway, Alex Schenk, Michael J. Sear, Jean-Philippe Tetienne, Alon Hoffman, Steven Prawer, Chris I. Pakes, Anton Tadich, Nathalie P. de Leon, Adam Gali, Lloyd C. L. Hollenberg

**Full NEXAFS Scans: *sp²* defect quantification**

NEXAFS measurements used an elliptically polarized undulator set to linear polarization in the range 270-340eV. During experiments the sample storage, annealing and measurement is conducted in a multi-chamber UHV system where all chambers are typically below $1\times10^{-9}$ mbar. XPS measurements are conducted to verify the cleanliness of the sample prior to measurement, using an hemispherical analyser. Beam steering control and sample targeting is assisted by a co-axial camera and fluorescent target crystal (YAG), while energy correction is conducted both with a reference gold foil (using XPS) and an amorphous carbon foil which is partially inserted into the beam during each individual NEXAFS scan. In cases where this energy reference scan did not provide high enough quality data for an accurate energy correction the diamond NEXAFS scan was manually adjusted such that the exciton peak positions of all data align (at 289.15 eV). This simultaneously aligns the second absolute band-gap dip for all samples (at ≈302.4 eV), which is the other typical method previously used for manual energy correction in diamond NEXAFS scans.

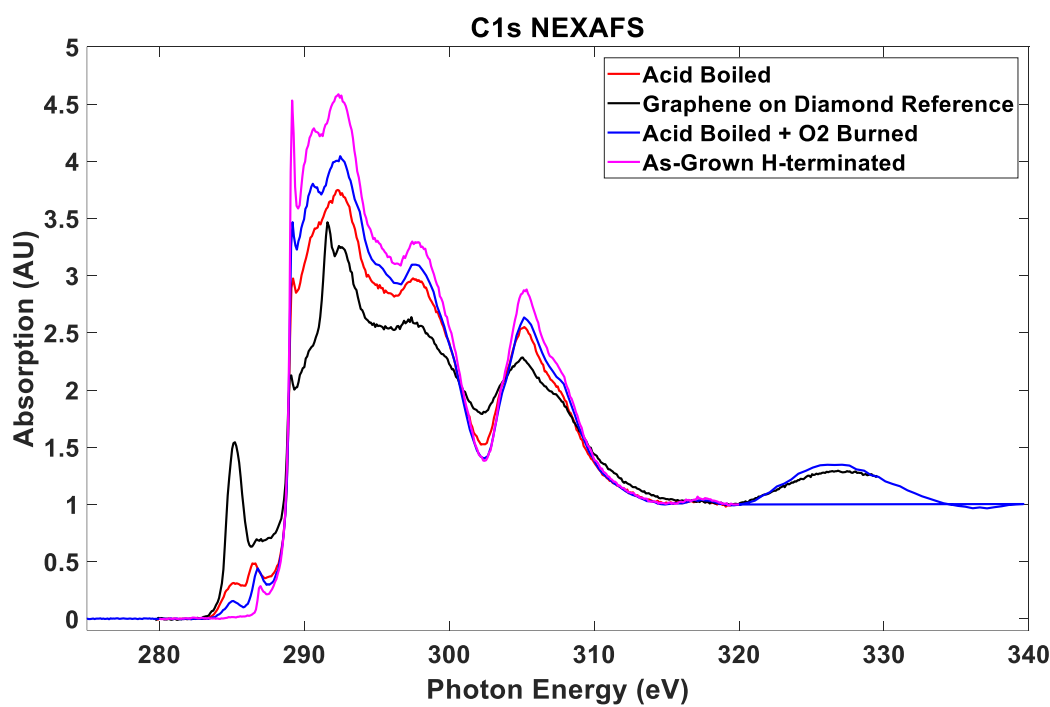

**Figure S1:** Full Partial-Electron Yield Carbon K-edge NEXAFS spectra for the samples shown in Figure 1.

While there are some minor observed differences in the oxygen chemistry for these surfaces, the pre-edge (band-gap <289eV) density of states differences between the samples in **Figure S1** are clearly dominated by the changes in *sp²* content, including significant changes in the C=C π* peak (285eV), and attendant changes to the C-C σ* profile, most easily seen as an inverse relationship between the π* peak height and the second absolute band-gap contrast (302.4eV). The diamond core-level exciton line (289.15eV) also shows an inverse relationship with C=C content, assumedly due to exciton relaxation effects, and the graphene σ* resonance can be observed at 291.7eV as expected.

From Fig S1 it can also be clearly seen that the monolayer graphene signal represents roughly 50% of the total NEXAFS signal intensity - most easily observed as a ≈1:1 ratio in the σ* ('bright line') onset step heights between the diamond (@289eV) and graphene (@291eV), and a reduction in the second absolute band-gap contrast by ≈2-3x). This is consistent with the 3.86Å mean-free path of 260eV C-KLL Auger electrons in diamond[1], and the fact that there is at least a monolayer of diamond termination (Oxygen/Hydrogen) atoms between the diamond and graphene.

Intercalated non-carbon species between the diamond and graphene could perceivably mask (reduce) the integrated σ* signal component from the diamond, artificially decreasing the measured $I_{ref}(\Delta E)$ value in equation (1) – repeated below.

$$f_{C=C} = \frac{I_{sam}^{\pi*}/I_{sam}(\Delta E)}{I_{ref}^{\pi*}/I_{ref}(\Delta E)} \qquad (1)$$

As such, the maximal decrease in $I_{ref}(\Delta E)$, and hence maximal decrease in calculated $f_{C=C}$ for each sample is a factor of 2x.

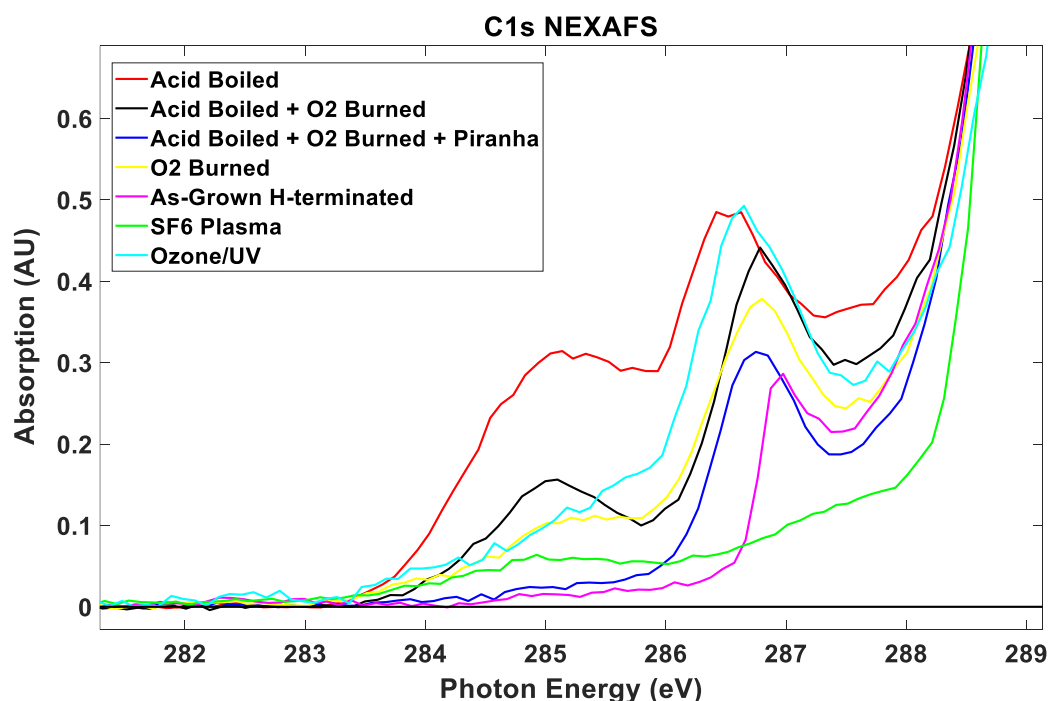

**Figure S2**: Pre-edge region of the Carbon K-edge NEXAFS of the various treated diamond surfaces, as mentioned in the main text.

## Bulk Fermi-level and Fermi-level Pinning

The bulk Fermi-level in the nitrogen-doped diamond sample is nominally set only by the donor level of the nitrogen (1.7eV), not considering other parasitic defects and acceptor levels. At a nitrogen concentration of $N_d = 1 \times 10^{15}/cm^3$ this results in a total electron carrier density of approximately $N_n = 5 \times 10^4/cm^3$, based on an intrinsic carrier density of $n_i \approx 10^{-27}/cm^3$ and equilibrium with thermally activated holes $N_p$, using the related equations:

$$N_p = n_i \times e^{\frac{(E_f - E_i)}{kT}} \tag{S1}$$

$$N_n = n_i \times e^{\frac{(E_i - E_f)}{kT}} + \frac{N_d}{1 + 2 \times e^{\frac{(E_f - E_d)}{kT}}} \tag{S2}$$

The Fermi-level pinning phenomenon can be directly observed in **Figure S3**, for a simulated sample with 0.1% surface coverage of *sp²* defects. Here a minimal change in the surface Fermi level is observed for a donor density up to a specific concentration, in this case about $10^{18}$ N/cm³:

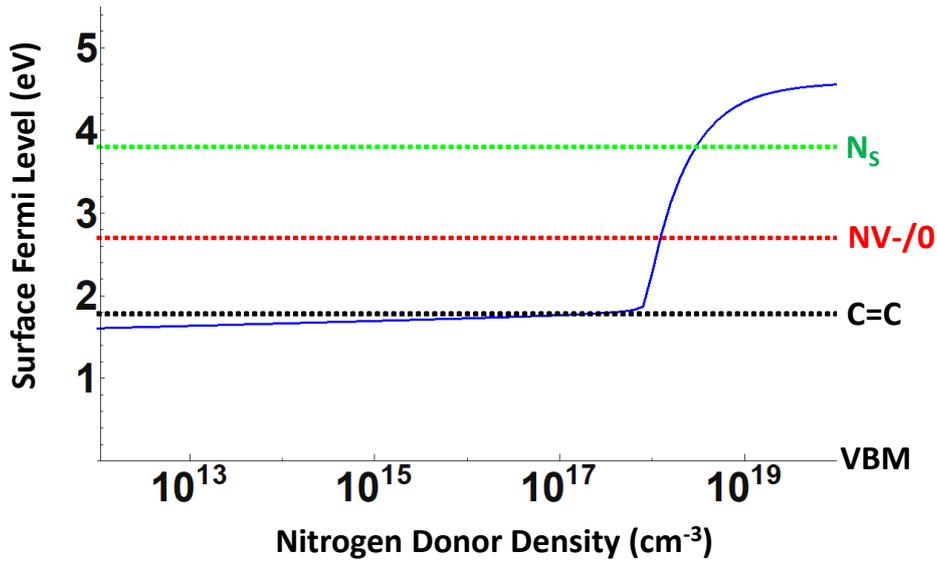

**Figure S3:** Surface Fermi-level as a function of nitrogen donor density (Nd in the main text). This is calculated for a surface *sp²* defect (C=C) concentration of 0.1%.

Changes in the fraction of *sp²* defects at the surface affect the donor concentration at which the Fermi-level pinning is overcome (**Figure S4**). In the case of N-implanted diamond, a 0.1% surface *sp²* concentration requires approximately $10^{18}$ N/cm³ nitrogen donors to avoid Fermi-level pinning. This equates to an implant of roughly $10^{12}$ N/cm², which generates an ensemble of NV centres under normal experimental conditions. Notwithstanding the need for increased understanding of photon-driven dynamic processes and effects of nanoscale inhomogeneity, this correlates well with our experience that the charge state (fluorescence) of ensemble NV samples are minimally perturbed by surface treatment processes and do show measurable NV centres with acid boiling (c.f. the single-NV sample presented in the main text).

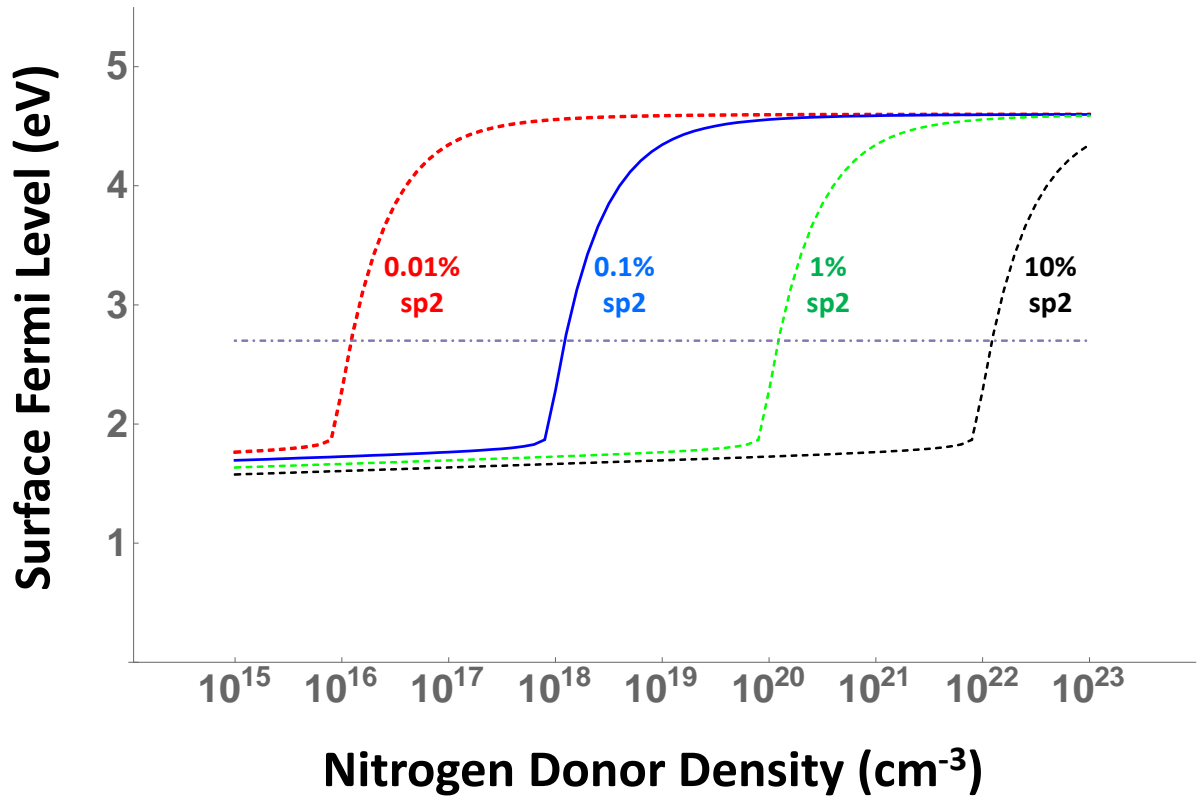

**Figure S4:** Surface Fermi-level as a function of nitrogen donor density, calculated for a range of surface *sp²* defect concentrations.

**Coherence and Depth Measurements of Near-Surface NV Centres**

After O2-burning of the NV-implanted sample, the Rabi contrast values were sufficient for spin coherence and depth measurements. The depth measurements utilized an NMR-based technique[2] for determining the depth of the NV centres, using the immersion oil as the hydrogen nuclear spin bath at the sample surface. The coherence measurements utilized a Carr-Purcell Meiboom-Gill (CPMG) 1024 pulse sequence to measure the decoupled $T_2$ coherence of each NV. As can be seen in **Figure S5**, the decoupled decoherence scales inversely with distance to the surface, while the Rabi contrast does not appear to be correlated with distance to the surface. As such, there is no observed correlation between the Rabi contrast and the decoupled coherence time of near-surface NV centres in this sample.

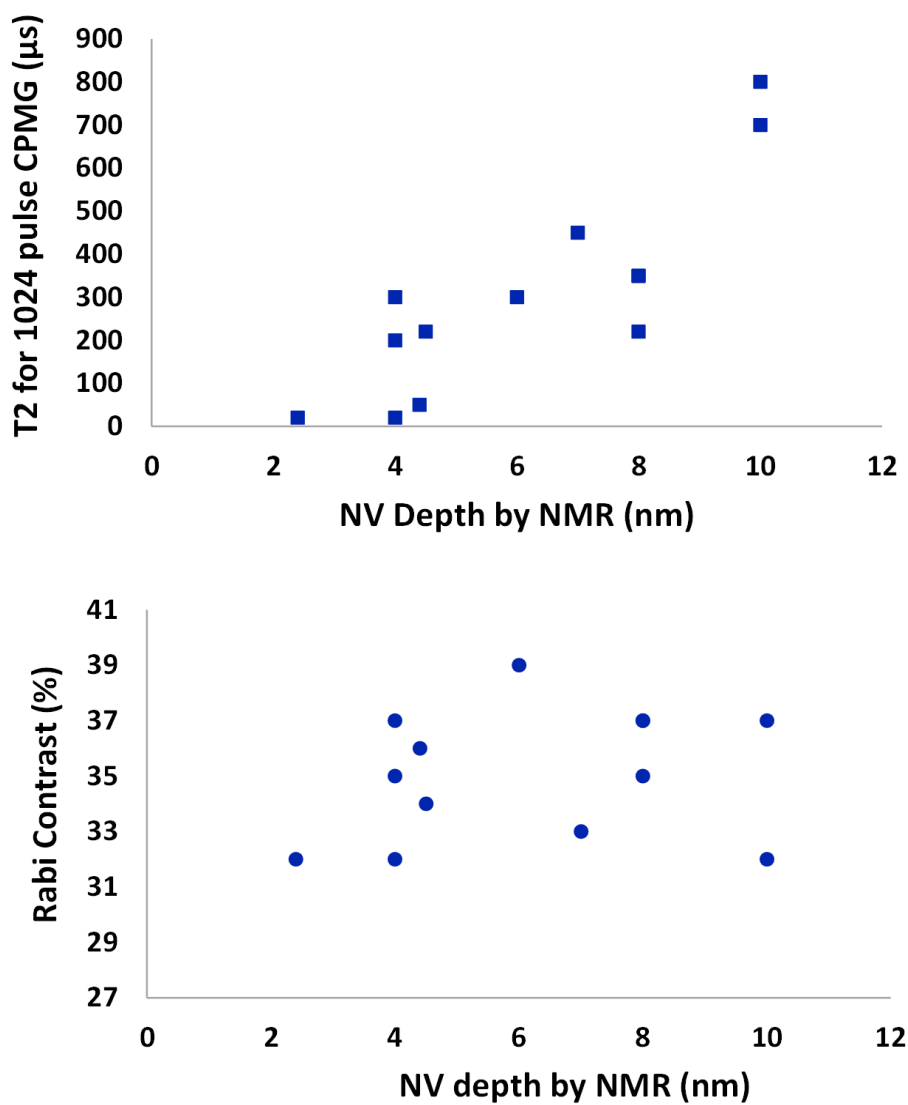

**Figure S5**: Decoupled coherence (T$_2$) values and Rabi contrast for NV centres measured in an O$_2$-burned sample, as a function of depth, found using NMR-based techniques.

**Computational methods**

The calculations were performed with the Vienna *Ab-initio* Simulation Package (VASP)[3]. The projector augmented-wave (PAW)[4] method was used to represent the electron-ionic core interactions. The energy cutoff for the expansion of plane-wave (charge density) is set to 370(740) eV. We used gradient-corrected Perdew-Burke-Ernzerhof (PBE) functional for exchange and correlation in the geometry optimization procedure. The electronic structure was determined by the screened hybrid functional HSE06[5] at the PBE geometries that is computationally demanding over PBE calculations but provides accurate band gap and surface bands[6]. The calculated lattice constant of diamond bulk is 3.57 Å which is in good agreement with experiment value of 3.567 Å. We used this optimized lattice constant to construct the a primitive C(100)-2×1 reconstructed surface, which was modelled by a repeating slab of fourteen carbon layers with a vacuum region more than 10 Å to avoid the interaction between the periodic images, as shown in **Fig. S6**(a). The 4×8×1 $\Gamma$-centered $k$-point mesh was applied to map the Brillouin zone. Each dangling bond on the surface is saturated by one hydrogen atom. The bottom two carbon layers and hydrogen atoms are fixed to the optimized reconstructed structure and the rest atoms are allowed to relax until the residual forces went below $2\times10^{-2}$ eV/Å. To study surface C=C $sp^2$ defect bonds, we build-up a 6×6 supercell with thickness of 14 layers from the optimized C(100)-2×1 primitive cell. The periodic lattice vectors in the $x$-$y$ plane were 15.15 Å×15.15 Å, thus $\Gamma$-point mesh is sufficient to obtain converged results. The band structures are calculated in the irreducible zone and the surface $sp^2$ states were verified from the site projected wave function character of each band.

We model the presence of the empty bands in NEXAFS spectrum by the adiabatic acceptor level of the corresponding surface defect. We used a two-step procedure to estimate this acceptor level in the diamond slab model. The vertical ionization energy of the defect was estimated by promoting an electron from the valence band maximum to the empty defect level. This approximation neglects the exciton binding energy which is in the order of few tens of meV. Then the relaxation energy of the acceptor state was calculated by the following processes: at the optimized geometry of neutral charged state, we calculate the total energy of the negatively charged state, then we applied the usual geometry optimization procedure to find the minimum total energy of the negatively charged state as a function of the coordinates of the ions. This provides the relaxation energy upon ionization. By subtracting the relaxation energy from the ionization energy, we obtain the adiabatic acceptor level. By this method, we avoid to compare the total energy of the neutral and charged supercells which would require a charge correction scheme with computationally demanding scaling methods[7].

**Convergence test on the slab size for the position of the empty levels**

We used PBE functional to assess the position of the empty levels in the bandgap as a function of the thickness of the slab. Firstly, we calculated the band structures of two typical cases: pristine H-terminated C(100) and H-terminated C(100) with two missing hydrogen atoms on a carbon dimer, as shown in Fig. 1(b). For a pristine H-terminated C(100) surface, the lowest unoccupied states are hydrogen images states located at ~2.95 eV above the valence band maximum (VBM). For a H-terminated C(100) surface with two missing hydrogen atoms on a carbon dimer, a surface C=C $sp^2$ state was found at $E_{VBM}$+1.97 eV in the bandgap, and the hydrogen image state was located at $E_{VBM}$+2.94 eV. Then, we calculated the energy difference as a function of layer thickness for these two cases, as shown in Fig. 1(c). Obviously, the energy difference appears to converge at thickness of 42 layers. We conclude that the empty levels are shifted up by 0.35 eV in the working slab model of C(100)-6×6-14L with respect to the converged slab model. We also examined the lateral size effect. Only 20 meV energy shift in the empty level was observed on a 12×12 large supercell indicated that a 6×6 supercell was sufficient to obtain converged position of the empty levels. Finally, we conclude that a -0.35 eV correction is required for the position of the empty level with respect to VBM in our working slab model.

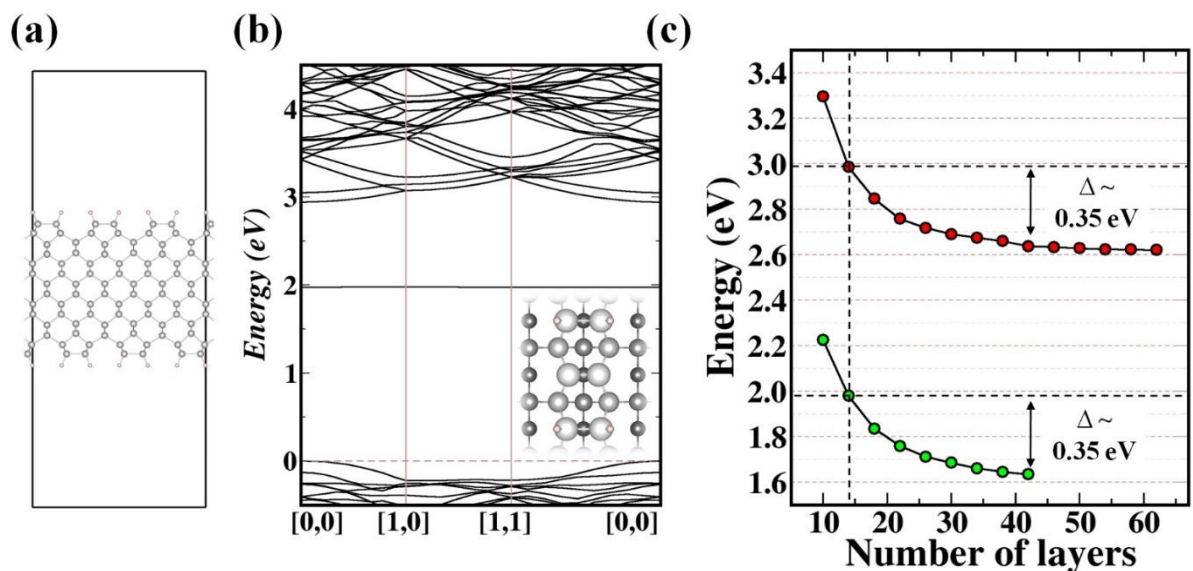

**Figure S6**. (a) The structure of diamond (100)-6×6-14L. (b) The band structure of H-terminated C(100)-6×6-14L surface with two missing hydrogen atoms on a carbon dimer forming a *sp²* C=C bond on the surface. The structure details are shown in the panel. The small pink balls are hydrogen atoms and the other balls are carbon atoms. The colour depth represents the height of carbon atoms that the topmost carbon atoms are white and the lowest lying carbon atoms are black. (c) The position of empty states (referred to VBM) as a function of slab thickness. The red balls are the results of pristine H-terminated C(100)-6×6 and the green balls are the results of H-terminated C(100)-6×6 with two missing hydrogen atoms.

**The *sp²* defect model on C(100) surface**

To investigate the different plausible structures for *sp²* defects on a C(100)-6×6-14L surface model, overall we have examined more than one hundred models (discussed in the next section). The negatively and positively charged states of the plausible configurations have also been examined. We calculated the band structure using PBE at the beginning, following by HSE calculations to obtain the precise position of *sp²* defect levels. The size effect (0.35 eV shift) and relaxation energy have also been considered to obtain the adiabatic acceptor level of the defect. Our results imply that the adiabatic acceptor level of this family of *sp²* defects appears at around 2.2 eV above the valence band maximum (VBM) of diamond for most of the defect configurations. Eventually, the *sp²* defect model with the closest acceptor level to the measured NEXAFS π* resonance is the one-vacancy configuration, as shown in **Fig. 7(a)**. The detailed lattice parameters are shown in Fig. 7(b). As can be seen, the lengths of C=C bonds are 1.39 Å which are apparently shorter than regular C-C bonds of 1.52 Å. The adiabatic acceptor level induced by this system's C=C bond is located at 1.78 eV above the VBM.

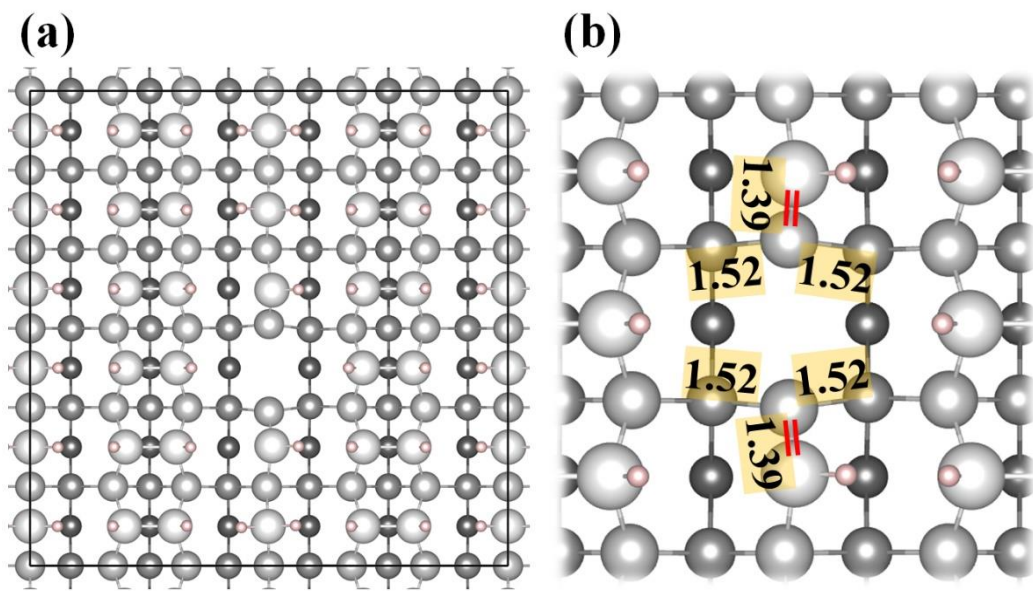

**Figure S7**. (a) The structure of one surface carbon vacancy model of diamond (100)-6×6. (b) The calculated geometries of one-surface-carbon-vacancy model. The bond lengths are shown in Ångström units. The position of C=C bonds are depicted by red double lines.

**Considered *sp²* defect configurations**

The considered *sp²* defect configurations are presented in this section. It is well-known that a regular *sp²* hybridized carbon bond length is apparently shorter than the standard *sp³* carbon bond length. From chemistry textbooks and databases, one can find the standard *sp²*-*sp²* C=C bond is ~1.34 Å and *sp³*-*sp³* C-C bond is ~1.54 Å. Such a formation of *sp²* carbon requires structure distortion or defects on the diamond surface. The most common defect on the surface would be the carbon vacancy. Therefore, we calculated numerous structures which can be classified into three types. Type 1 is pristine diamond surface without carbon vacancy defect, as shown in **Fig. S8**. Missing hydrogen (e.g., Fig. S8-i) or carbon dimer distortion (e.g., Fig. S8-ii) can form a carbon-carbon bond shorter than 1.5 Å. Type 2, as shown in **Fig. S9**, is the configuration with one missing carbon. This carbon vacancy could be placed at the first layer (e.g. Fig. S9-i), the second layer (e.g. Fig. S9-xx), or the third layer (e.g. Fig. S9-xxi). We also calculated single carbon vacancy configuration on the step edge which is not presented here. Type 3, as shown in **Fig. S10**, is multi-vacancy configuration. The possible combination configurations and charged defects were also calculated.

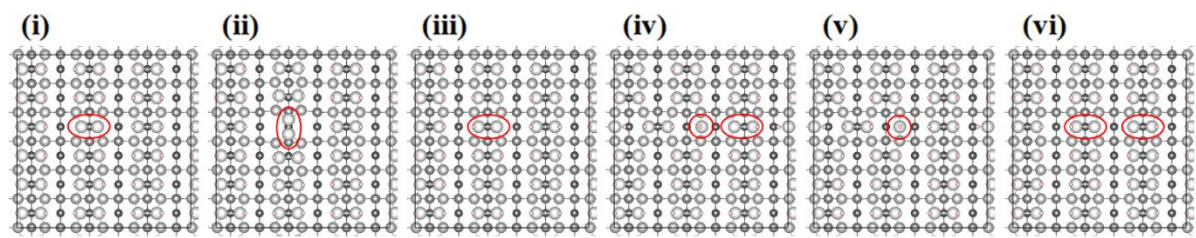

**Figure S8**. Considered configurations of *sp²* carbon on the pristine diamond surface. The position of *sp²*-*sp²* bonds are highlighted with red circles.

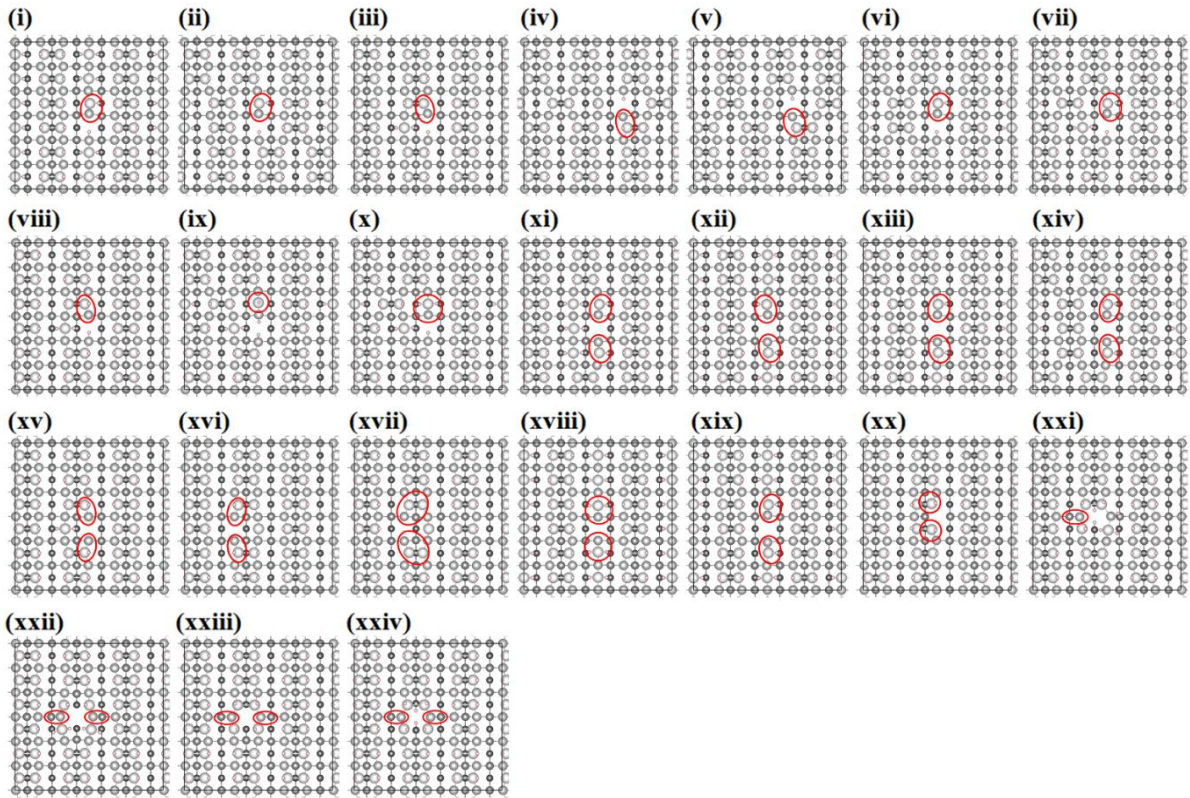

**Figure S9.** Considered configurations of *sp*$^2$ carbon with single carbon vacancy on the diamond surface. The position of *sp*$^2$-*sp*$^2$ bonds are highlighted with red circles.

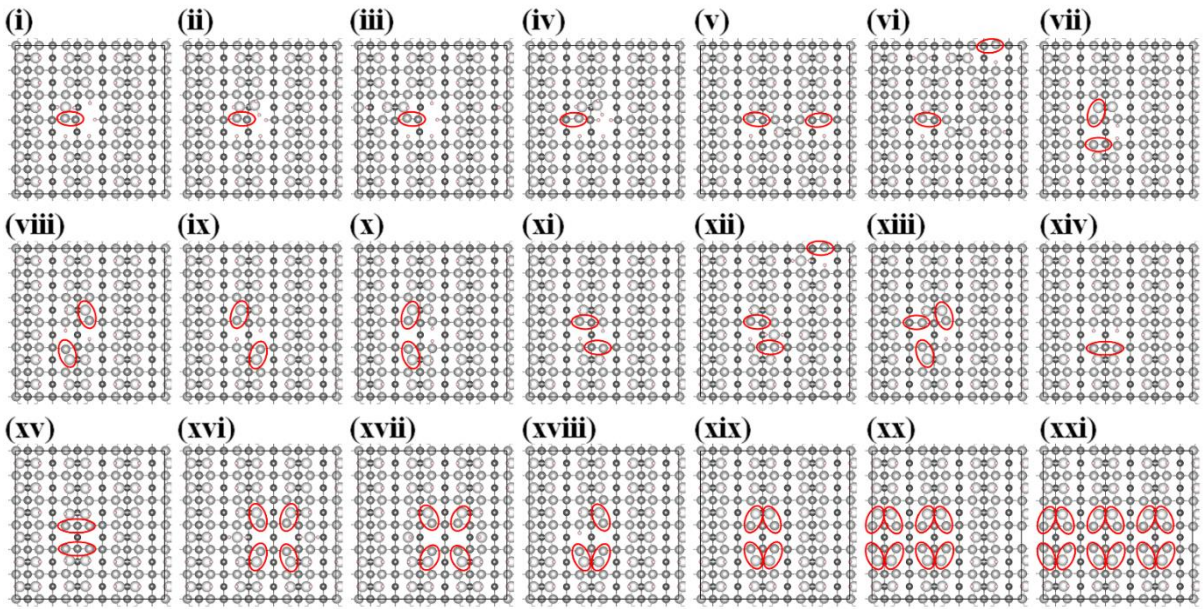

**Figure S10.** Considered configurations of *sp*$^2$ carbon with multi-vacancies on the diamond surface. The position of *sp*$^2$-*sp*$^2$ bonds are highlighted with red circles.

We calculated the electronic structure of these selected configurations at PBE level. To obtain the correct position of empty state, we include three corrections: (1) thickness effect which shifts *downward* the empty states around 0.35 eV; (2) HSE06 quasi-particle level correction which shift *upward* the empty states by around 0.9 eV; (3) relaxation energy which shifts *downward* the adiabatic acceptor level roughly by 0.3 eV (see exceptions below). Therefore, the calculated PBE empty defect level would lead to a corrected adiabatic acceptor level that lies by about 0.3 eV above the PBE value for most of the configurations. This rule of thumb can be used to screen almost all the defect configurations at PBE level. We found that single *sp$^2$*-*sp$^2$* bond on diamond surface with either single carbon vacancy (e.g., Fig. S8, i-x) or multiple vacancies (e.g., Fig. S9, i-iv), possess adiabatic acceptor levels always above $E_{VBM}$+2.0 eV that does not explain the experimental NEXAFS features. Configurations with three *sp$^2$*-*sp$^2$* bonds (e.g., Fig. S10 xiii) also results in empty levels above $E_{VBM}$+2.0 eV. It is worthy to mention that for some cases (e.g., Fig. S9-xix), we found an empty defect level below $E_{VBM}$+1.3 eV by PBE calculation. However, the relaxation energy in this case is unexpectedly large, 0.9 eV, thus after energy corrections the adiabatic acceptor level eventually drops down to $E_{VBM}$+0.9 eV. It cannot be excluded that the vibration assisted excitation of this configuration contributes to the NEXAFS spectrum. In general, only the configurations of two single *sp$^2$*-*sp$^2$* bonds (e.g., Fig. S9, xi-xvi) exhibit an adiabatic acceptor level at <$E_{VBM}$+2.0 eV, in particular, the configuration of Fig. S9-xi has the lowest adiabatic acceptor level at $E_{VBM}$+1.78 eV which is in broad agreement with the experimental findings.